\documentclass[aps,showpacs,preprintnumbers,amsmath,amssymb]{revtex4}
\usepackage{dcolumn}
\usepackage[dvips]{epsfig}

\usepackage{float}
\usepackage{graphics,epsfig}
\usepackage{graphicx}
\usepackage{dcolumn}
\usepackage{bm}

\begin{document}
\baselineskip=0.8 cm
\title{\bf Strong Gravitational Lensing of Quasi-Kerr Compact Object with Arbitrary Quadrupole Moments  }

\author{
Changqing Liu$^{1,2}$, 
Songbai Chen$^{1}$, 
and Jiliang  Jing$^{1}$\footnote{Corresponding author, Email:
jljing@hunnu.edu.cn} 
}

\affiliation{1) Department of Physics, and Key Laboratory of Low Dimensional Quantum Structures and Quantum Control of Ministry of Education, Hunan Normal University,  Changsha, Hunan 410081, P. R. China}

\affiliation{2) Department of Physics and Information Engineering, Hunan Institute of Humanities Science and Technology, Loudi, Hunan 417000, P. R. China}

\begin{abstract}
\baselineskip=0.6 cm
\begin{center}
{\bf Abstract}
\end{center}

We study the strong gravitational lensing on the equatorial plane of a quasi-Kerr
 compact object with arbitrary quadrupole moments which can be used to model
 the super-massive central object of the galaxy. We find that,
 when the quadrupolar correction parameter $\xi$ takes the positive (negative) value,
 the photon-sphere radius $r_{ps}$, the minimum impact parameter $u_{ps}$, the coefficient $\bar{b}$,
 the relative magnitudes $r_m$ and the angular position of the relativistic
 images $\theta_{\infty}$ are larger (smaller)  than the results obtained
 in the Kerr black hole, but the coefficient $\bar{a}$, the deflection
 angle $\alpha(\theta)$ and the angular separation $s$ are smaller (larger) than
 that in the Kerr black hole.    These features may offer a way to probe
 special properties for some rotating compact objects by the astronomical instruments in the future.

\end{abstract}

 \maketitle
\newpage
\section{Introduction}

In the framework of general relativity, the no-hair theorem \cite{noh} guarantees that a neutral rotating astrophysical black hole is uniquely described by the Kerr metric only with two parameters, the mass $M$ and the rotational parameter $a$. Observations of the weak gravitational systems agree well with the prediction of the general relativity. However, the hypothesis that the astrophysical black-hole candidates are described by the
Kerr metric still lacks the direct evidence, and the
general relativity has been tested only for weak gravitational
fields. In the regime of strong gravity, the general relativity
could be broken down and astrophysical black holes might not be the Kerr
black holes as predicted by the no-hair theorem \cite{FCa,TJo,CBa,psa08}.
Several parametric deviations from the Kerr metric, i.e., multipole moments, have been
suggested to study observational signatures  including
gravitational waves from extreme mass-ratio inspiral (EMRI)
\cite{gwave,ga1,ga2,JGa,Apostolatos} and the electromagnetic spectrum emitted by
the accreting disk around black holes \cite{ag1,CBa3}.

The general stationary axisymmetric  neutral compact object can been described in terms of its mass, rotational parameter and multipole moments \cite{moments}.
The multipole moments are consist of a set of mass multipole moment $M_l$ and current multipole moment $S_l$, here the subscript $l$ of them is labeled by the angular inter eigenvalue $l\geq0$. The relation between the parameters of the multipole moments can be expressed as \cite{moments}
\begin{eqnarray}
M_l + {\rm i}S_l = M({\rm i}a)^l + \delta M_l + {\rm i}\delta S_l.
\label{deltamult}
\end{eqnarray}
For the Kerr black hole, the deviation $\delta M_l$ and $\delta S_l$ are equal to zero. Thus the relation (\ref{deltamult}) is unique to the Kerr black hole and is just a mathematical expression of the famous no-hair
theorem. Therefore, for a general rotating spacetime, by measuring three independent multipole moments of the spacetime we could provide sufficient information for verifying whether or not the central body is a Kerr black hole. Now, four approaches for measuring the multipole moments or a perturbation of the Kerr black hole have been proposed to test the no-hair theorem. The first approach based on writing the general stationary axisymmetric metric in terms of multipole moments
was proposed by Ryan and was extended by Barack \textit{et al}.\cite{gwave}. They demonstrated how the laser interferometer space antenna (LISA) detector could map compact object deviating from the Kerr metric by means of EMRI observations. In the second approach, Collins and Hughes \cite{ga1} introduced ``bumpy black hole" (a spacetime that slightly deviates from the exact black hole of general relativity). This approach was recently generalized to the case of a rotating black hole in an alternate theories of gravity \cite{ga2,ag4}.
In the third one,  based on the Manko-Novikov metric \cite{ag3}, Apostolatos \textit{et al} \cite{Apostolatos} showed that the appearance of Birkhoff chains in the neighborhood of a resonant tori would lead to such a modification of a gravitational-wave measurement.
Finally, in the fourth one, Glampedakis \textit{et al} \cite{Glampedakis} proposed a quasi-Kerr metric by choosing the (dimensionless) quadrupole moment to be $q_{\rm Kerr}-\xi$, where the quadrupolar correction  parameter $\xi$  represents a potential deviation from the Kerr metric and $q_{\rm Kerr}=-J^2/M^4=-a^2/M^2$. And the quadrupole moment \cite{Glampedakis} can be written as
\begin{equation}
Q=-M\left(a^2+\xi M^2\right).
\label{qradmoment}
\end{equation}
Glampedakis \textit{et al} calculated the periastron precession and constructed `kludge' gravitational waveforms as a function of the parameter $\xi$. These waveforms can be significantly different from the expected Kerr signal even for small changes of the quadrupole moment. In a series of papers \cite{Johannsen}, Johannsen \textit{et al} analyzed in detail the
innermost stable circular orbit, the circular photon orbit and the electromagnetic spectrum of the quasi-Kerr spacetime to test the no-hair theorem.

Gravitational lensing caused by deflection of light rays in a gravitational field is an ordinary phenomenon in astronomical observations when the light pass through the massive compact objects such as black
hole, quasars and supernova.  In the last decades, the theory of gravitational lensing has been developed along two branches. The former takes the weak deflection limit as photon radius is much larger than the gravitational radius of the lens. The latter uses the strong deflection limit as the light ray loops around the massive compact object many times before it reaches to the observer. By this mechanism, two infinite series of relativistic images appears on each side of the lens. These relativistic images can provide us not only some
important signatures about compact objects in the universe, but also profound verification of alternative theories of gravity in the strong field regime   \cite{Vir,Vir1,Vir2,Vir3, Bozza2,Bozza3}. Thus,
the strong gravitational lensing is regarded as a powerful indicator of the physical nature of the central celestial objects and then has been studied extensively in various theories of gravity
\cite{Gyulchev,Gyulchev1,Fritt,Bozza1,Eirc1,whisk,Bhad1,Song1,Song2,TSa1,AnAv,
Ls1,Darwin}.

The main purpose of this paper is to study the strong gravitational lensing by the quasi-Kerr compact object \cite{Glampedakis} and to see whether it
can leave us the signature of the quadrupole moment parameter in the photon sphere radius, the deflection angle, the coefficients and the observable quantities of strong gravitational lensing. Moreover, we will explore how it differs from the Kerr black hole lensing.

The paper is organized as follows: In Sec. II, we will review briefly the metric of the quasi-Kerr compact object with the quadrupole moment parameter proposed by
Glampedakis \textit{et al} \cite{Glampedakis}  and calculate the radius of photon sphere. In Sec. III, we study the physical properties of the strong gravitational lensing by  quasi-Kerr compact object and probe the effects of the quadrupole moment parameter on the deflection angle, the coefficients and the observable quantities for gravitational lensing in the strong field limit. We end the paper with a summary.

\section{Rotating quasi-Kerr spacetime and radius of photon sphere}

With the help of the Hartle-Thorne spacetime \cite{HT}, the rotating quasi-Kerr compact object with arbitrary quadrupole moments was proposed by Glampedakis \textit{et al} \cite{Glampedakis}. This metric has three independent parameters, i.e., the mass $M$, the spin $a$ and the quadrupolar correction parameter $\xi$. The parameter $\xi$ describes the deviation of the quadrupole moments from the Kerr value.  The Kerr metric $g_{ab}^{K}$  in the standard Boyer-Lindquist coordinates can be expressed as
\begin{eqnarray}\label{kerr}
ds^2=-\Big(1-\frac{2Mr}{\Sigma}\Big)~dt^2-
\Big(\frac{4Mar\sin^2\theta}
{\Sigma}\Big)~dtd\phi
+\frac{\Sigma}{\Delta}~dr^2+\Sigma~d\theta^2+
\Big(r^2+a^2+\frac{2Ma^2r\sin^2\theta}{\Sigma}\Big)
\sin^2\theta~d\phi^2
\end{eqnarray}
with
\begin{eqnarray}
\Delta\equiv r^2-2Mr+a^2,~~~
\Sigma\equiv r^2+a^2\cos^2~\theta.
\end{eqnarray}
The quadrupolar correction is introduced by choosing a quadrupole moment of the form (\ref{qradmoment}) in the Hartle-Thorne metric. Then, the quasi-Kerr metric $g_{ab}^{QK}$ in the Boyer-Lindquist coordinates is given by  \cite{Glampedakis}
\begin{equation}
g_{ab}^{QK}=g_{ab}^{K}+\xi h_{ab}+{\cal O}(\delta M_{\ell \geq 4}, \delta S_{\ell \geq 3}),
\label{qKerr}
\end{equation}
\noindent
with
\begin{eqnarray}
h^{\rm tt}&=&(1-2M/r)^{-1}\left[\left(1-3\cos^2\theta\right)
\mathcal{F}_1(r)\right],\nonumber \\
h^{\rm rr}&=&(1-2M/r)\left[\left(1-3\cos^2\theta\right)
\mathcal{F}_1(r)\right],
\nonumber \\
h^{\rm \theta\theta}&=&-\frac{1}{r^2}\left[\left(1-3\cos^2
\theta\right)\mathcal{F}_2(r)\right],
\nonumber \\
h^{\rm \phi\phi}&=&-\frac{1}{r^2\sin^2\theta}\left[\left(1-3
\cos^2\theta\right)\mathcal{F}_2(r)\right], \nonumber \\
h^{\rm t\phi}&=&0,
\end{eqnarray}
where the functions $\mathcal{F}_{1}(r)$ and $\mathcal{F}_{2}(r)$ are given in $appendix A$ of the Ref. \cite{Glampedakis}.
The radius of the event horizon of the quasi-Kerr black hole, obtained from the relation $ g_{t\phi}^2-g_{tt}g_{\phi} g_{\phi}=0$, increases with increasing positive values of the quadrupolar correction parameter $\xi$ but decreases as the spin $a$ increases \cite{Johannsen}. However, if $\xi$ is negative, the event horizon is absent and the quasi-Kerr spacetime exists a naked singularity.

\begin{figure}[ht]
\begin{center}
\includegraphics[scale=1.1]{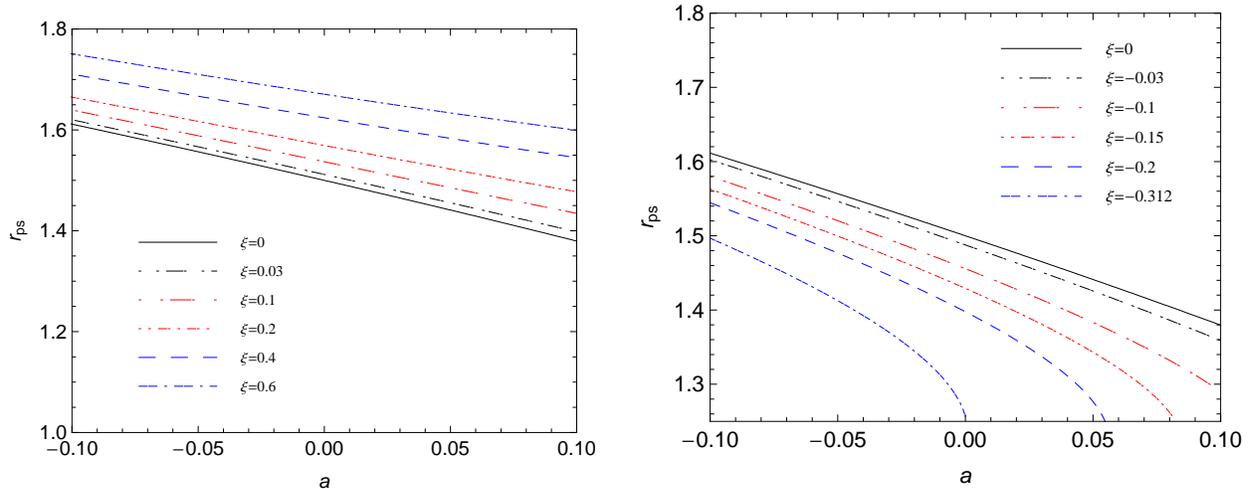}
\caption{Variation of the radius of the photon sphere with  the parameter $a$ for different quadrupolar
correction parameter $\xi$ in the quasi-Kerr spacetime.}\label{figure1}
\end{center}
\end{figure}

Let us now study the strong gravitational lensing of the rotating non-Kerr compact object. As in
refs. \cite{Vir,Vir1,Vir2,Vir3,Bozza2,Bozza3, Gyulchev,Gyulchev1,Fritt,Bozza1,Eirc1,
whisk,Bhad1,Song1,Song2,TSa1,AnAv,Ls1}, we just consider that both the observer and the source lie in the equatorial plane of the rotating quasi-Kerr compact object and the whole trajectory of the photon is limited on the same plane. Using the condition $\theta=\pi/2$ and taking $2M=1$, the metric (\ref{qKerr}) is reduced to
\begin{eqnarray}
ds^2&=&-A(r)dt^2+B(r)dr^2+C(r) d\phi^2-2D(r)dtd\phi, \label{metric1}
\end{eqnarray}
with
\begin{eqnarray}
A(r)&=&1-\frac{1}{r}+\frac{5\xi}{16r^5}\left[2a^2(6r^2+3r-1)+r^3(12r^3-18r^2+4r+1)\right]
\\\nonumber&-&\frac{5\xi}{16r^5}[6(2(r-1)^2r^5+a^2r(2r^2-1))\ln(\frac{r}{r-1})],\\
B(r)&=&\frac{r^2}{a^2-r-r^2}+\frac{5\xi~r^2[12r^3+18r^2-
4r-1+12(r-1)^2r^2\ln(\frac{r}{r-1})]}{16(a^2+(r-1)r)^2},\\
C(r)&=&a^2+\frac{a^2}{r}+r^2+\frac{5\xi~a^2[12r^3+18r^2-
4r-1+12(r-1)^2r^2\ln(\frac{r}{r-1})]}{16(r-1)^2r^2}
\\\nonumber&+&\frac{5\xi[(r^3+a^2(1+r))^2(6r^2-3r+1+(6r^3-3r)\ln(\frac{r}{r-1}))]}{8r^5},\\
D(r)&=&\frac{a}{r}+\frac{5\xi~a[r^3(-12r^2+12r-1)-2a^2(6r^4+3r^3-7r^2-3r+1)]}{16(r-1)r^5}
\\\nonumber&+&\frac{5\xi~a[6(r^4(2r^2-3r+1)+a^2(2r^5-3r^3+r))\ln(\frac{r}{r-1})]}{16(r-1)r^5}
.
\end{eqnarray}
Then, the null geodesics for the metric (\ref{metric1}) can be expressed as
\begin{eqnarray}
\frac{dt}{d\lambda}&=&\frac{C(r)-LD(r)}{D(r)^2+A(r)C(r)},\label{u3}\\
\frac{d\phi}{d\lambda}&=&\frac{D(r)+LA(r)}{D(r)^2+A(r)C(r)},\label{u4}\\
\bigg(\frac{dr}{d\lambda}\bigg)^2&=&\frac{C(r)-2LD(r)-L^2A(r)}{B(r)C(r)[D(r)^2+A(r)C(r)]},
\end{eqnarray}
where $\lambda$ is an affine parameter along the geodesics and $L$
is the angular momentum of the photon. With the condition $\frac{dr}{d\lambda}|_{r=r_0}=0 $, we can obtain the impact parameter $u(r_0)$
\begin{eqnarray}
u(r_0)=L(r_0)=\frac{-D(r_0)+\sqrt{A(r_0)C(r_0)+D^2(r_0)}}{A(r_0)}.
\end{eqnarray}
Moreover, the photon sphere is a time-like hyper-surface ($r=r_{ps}$)
on which the deflect angle of the light becomes unboundedly large as
$r_0$ tends to $r_{ps}$. In this spacetime, the equation for the photon sphere
reads
\begin{eqnarray}\label{rpsss}
A(r)C'(r)-A'(r)C(r)+2L[A'(r)D(r)-A(r)D'(r)]=0.\label{root0}
\end{eqnarray}
The biggest real root external to the horizon of this equation is
defined as the radius of the photon sphere $r_{ps}$.
Obviously, this equation is more complex than that in the background
of a static and Kerr black holes \cite{Vir2}. Therefore, it is impossible to get an analytical form for the photon sphere radius in this case. In Fig. \ref{figure1}, we present the
variety of the photon-sphere radius $r_{ps}$ with the rotational parameter $a$ for different quadrupolar
correction parameter $\xi$ numerically. The figure shows us that the photon sphere radius $r_{ps}$ decreases as the rotational parameter $a$ increases, but it increases as the quadrupolar correction parameter $\xi$ increases. Moreover, we also find that the photon sphere radius $r_{ps}$ exists
only in the regime $\xi>-0.312$ when the quasi-Kerr compact object rotates in the same direction as the photon ($a>0$). As the quadrupolar correction parameter $\xi$ becomes negative the radius of the photon sphere becomes smaller,  which implies
that the photons are more easily captured by the quasi-Kerr compact object with the negative quadrupolar correction parameter $\xi$ than that of the Kerr black
hole.

\section{Strong gravitational lensing of rotating quasi-Kerr compact object}

In this section we will study the gravitational lensing of the
rotating quasi-Kerr compact object which has a photon sphere and then
probe the effects of the quadrupolar correction parameter $\xi$ on the
coefficients and the observable quantities of the strong gravitational lensing.

\subsection{Coefficients of strong gravitational lensing}

The deflection angle for the photon coming from infinite in a
stationary, axially-symmetric spacetime, described by the metric
(\ref{metric1}) obeys \cite{Ein1}
\begin{eqnarray}
\alpha(r_{0})=I(r_{0})-\pi,
\end{eqnarray}
with
\begin{eqnarray}
I(r_0)=2\int^{\infty}_{r_0}\frac{\sqrt{B(r)|A(r_0)|}[D(r)+LA(r)]dr}{\sqrt{D^2(r)+A(r)C(r)}
\sqrt{sgn(A(r_0))[A(r_0)C(r)-A(r)C(r_0)+2L[A(r)D(r_0)-A(r_0)D(r)]]}},\label{int1}
\end{eqnarray}
where $sgn(X)$ gives the sign of $X$.

It is obvious that the deflection angle increases as the parameter
$r_0$ decreases. For a certain value of $r_0$ the deflection angle
becomes $2\pi$, so that the light ray makes a complete loop around
the lens before reaching the observer. If $r_0$ is equal to the
radius of the photon sphere $r_{ps}$, we can find that the
deflection angle diverges and the photon is captured by the compact object.

In order to find the behavior of the deflection angle when the photon is close to the photon sphere, we use the evaluation method proposed by Bozza \cite{Bozza2}. The divergent
integral in Eq. (\ref{int1}) is first split into the divergent part
$I_D(r_0)$ and the regular one $I_R(r_0)$, and then both of them are
expanded around $r_0=r_{ps}$ with sufficient accuracy. This technique has been widely used in the study of the strong gravitational lensing for various
black holes
\cite{Vir,Vir1,Vir2,Vir3,Bozza2,Bozza3,Gyulchev,Gyulchev1,Fritt,Bozza1,Eirc1,
whisk,Bhad1,Song1,Song2,TSa1,AnAv,Ls1,Darwin}. Let us now to define
a variable
\begin{figure}[ht]
\begin{center}
\includegraphics[scale=1.1]{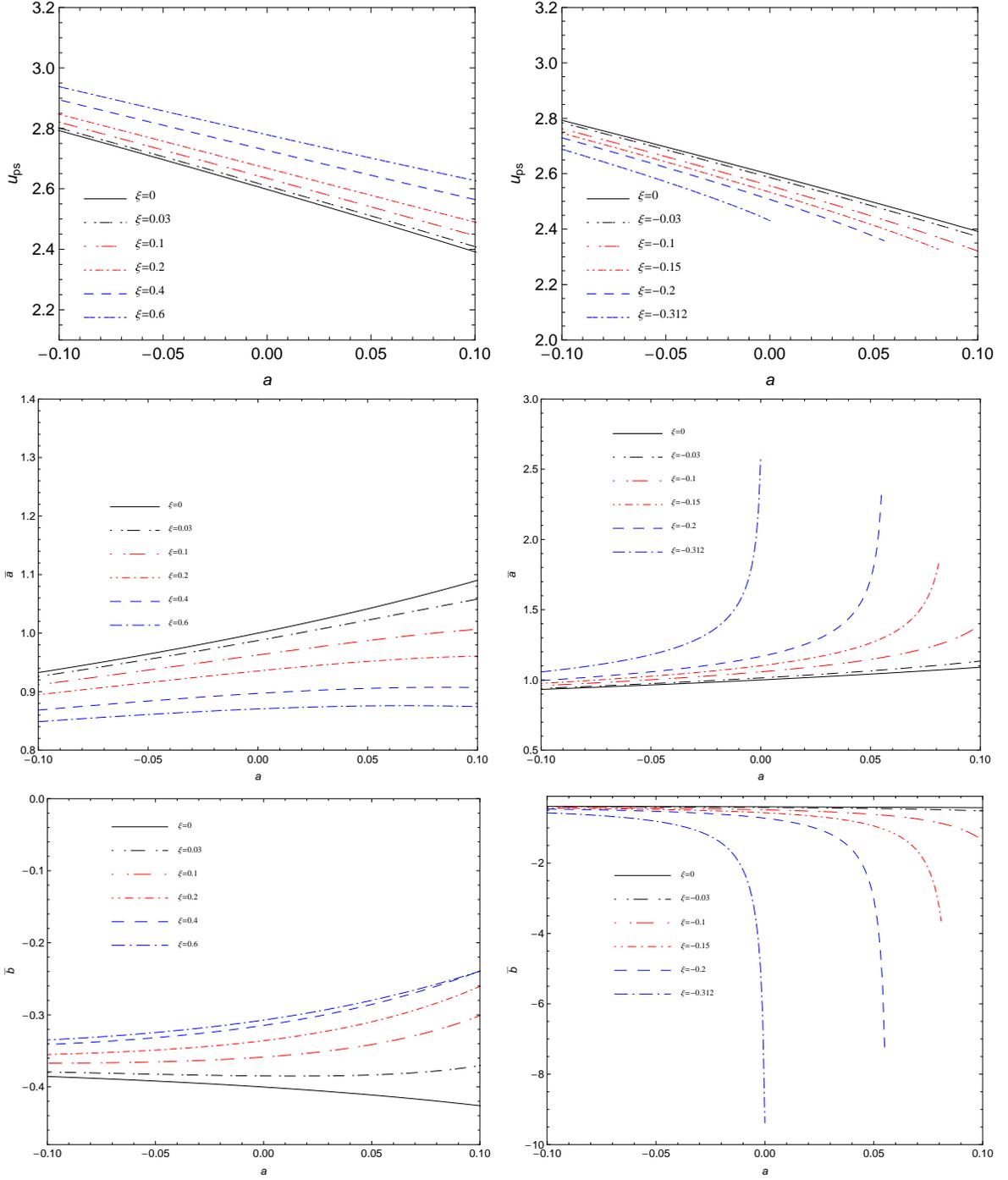}\caption{Variation of the coefficients for the
strong gravitational lensing with the parameter $a$ for different quadrupolar
correction parameter $\xi$ in the rotating quasi-Kerr spacetime.}
\label{ coefficients}
\end{center}
\end{figure}
\begin{figure}[ht]
\begin{center}
\includegraphics[scale=1.0]{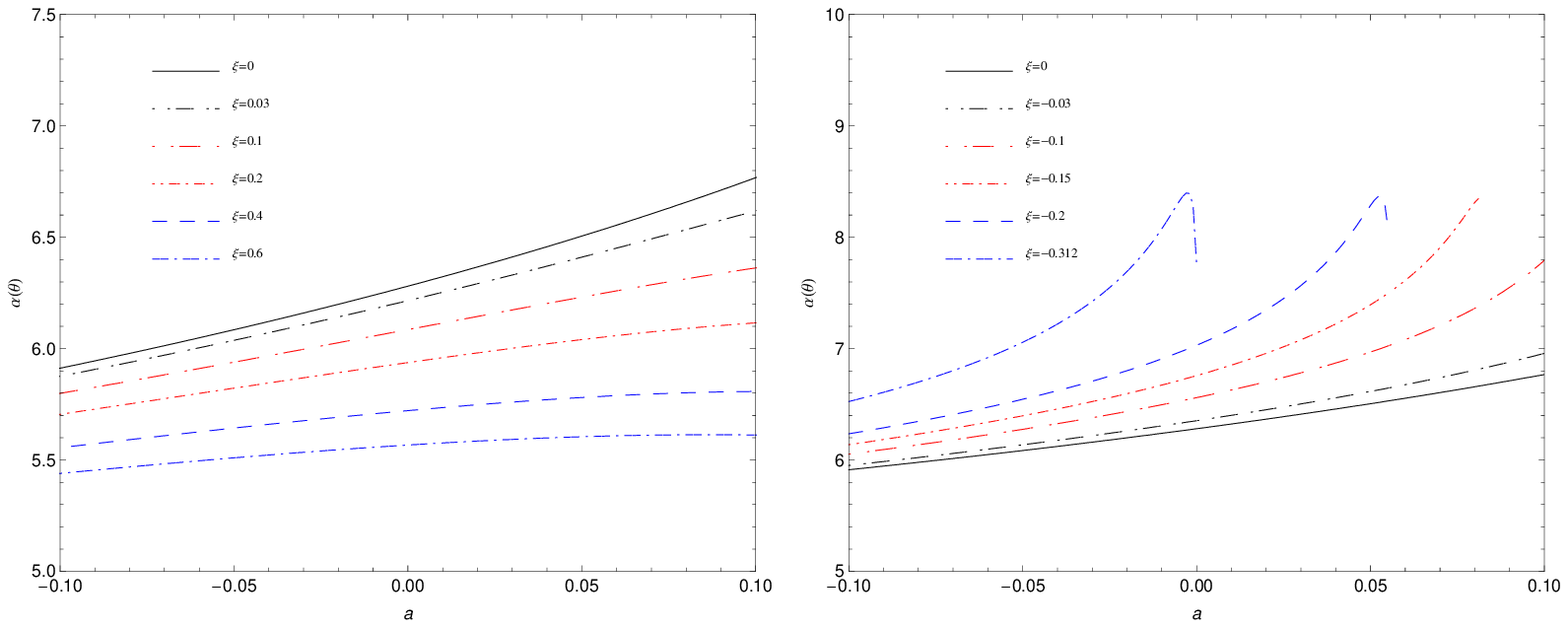}\caption{Variation of deflection angles evaluated at $u=u_{ps}+0.00326$ with the parameter $a$ for different quadrupolar
correction parameter $\xi$ in the rotating quasi-Kerr spacetime.}\label{figure3}
\end{center}
\end{figure}
\begin{eqnarray}
z=1-\frac{r_0}{r},
\end{eqnarray}
and rewrite the Eq.(\ref{int1}) as
\begin{eqnarray}
I(r_0)=\int^{1}_{0}R(z,r_0)f(z,r_0)dz,\label{in1}
\end{eqnarray}
with
\begin{eqnarray}
R(z,r_0)&=&\frac{2r_0}{\sqrt{C(z)}(1-z)^2}\frac{\sqrt{B(z)|A(r_0)|}[D(z)+LA(z)]}{\sqrt{D^2(z)+A(z)C(z)}},
\end{eqnarray}
\begin{eqnarray}
f(z,r_0)&=&\frac{1}{\sqrt{sign(A(r_0))[A(r_0)-A(z)\frac{C(r_0)}{C(z)}+\frac{2L}{C(z)}(A(z)D(r_0)-A(r_0)D(z))]}}.
\end{eqnarray}
Obviously, the function $R(z, r_0)$ is regular for all values of $z$
and $r_0$. However, the function $f(z, r_0)$ diverges as $z$ tends
to zero, i.e., as the photon approaches the photon sphere. Thus, the
integral (\ref{in1}) can be separated into two parts $I_D(r_0)$ and
$I_R(r_0)$
\begin{eqnarray}
I_D(r_0)&=&\int^{1}_{0}R(0,r_{ps})f_0(z,r_0)dz, \nonumber\\
I_R(r_0)&=&\int^{1}_{0}[R(z,r_0)f(z,r_0)-R(0,r_0)f_0(z,r_0)]dz
\label{intbr}.
\end{eqnarray}
Expanding the argument of the square root in $f(z,r_0)$ to the
second order in $z$, we have
\begin{eqnarray}
f_0(z,r_0)=\frac{1}{\sqrt{p(r_0)z+q(r_0)z^2}},
\end{eqnarray}
where
\begin{eqnarray}
p(r_0)&=&\frac{r_0}{C(r_0)}\bigg\{A(r_0)C'(r_0)-A'(r_0)C(r_0)+2L[A'(r_0)D(r_0)-A(r_0)D'(r_0)]\bigg\},  \nonumber\\
q(r_0)&=&\frac{r_0}{2C^2(r_0)}\bigg\{2\bigg(C(r_0)-r_0C'(r_0)\bigg)\bigg([A(r_0)C'(r_0)-A'(r_0)C(r_0)]
+2L[A'(r_0)D(r_0)-A(r_0)D'(r_0)]\bigg)\nonumber\\&&+r_0C(r_0)
\bigg([A(r_0)C''(r_0)-A''(r_0)C(r_0)]+2L[A''(r_0)D(r_0)-A(r_0)D''(r_0)]\bigg)\bigg\}.\label{al0}
\end{eqnarray}
From Eq.(\ref{al0}), we can find that if
$r_{0}$ approaches the radius of photon sphere $r_{ps}$ the
coefficient $p(r_{0})$ vanishes and the leading term of the
divergence in $f_0(z,r_{0})$ is $z^{-1}$, which implies that the
integral (\ref{in1}) diverges logarithmically. The coefficient
$q(r_0)$ takes the form
\begin{eqnarray}
q(r_{ps})&=&\frac{sgn(A(r_{ps})) r^2_{ps}}{2C(r_{ps})}\bigg\{A(r_{ps})C''(r_{ps})-A''(r_{ps})C(r_{ps})+
2L[A''(r_{ps})D(r_{ps})-A(r_{ps})D''(r_{ps})]\bigg\}.
\end{eqnarray}
Therefore, the deflection angle in the strong field region can be
expressed as \cite{Bozza2}
\begin{eqnarray}
\alpha(\theta)=-\bar{a}\log{\bigg(\frac{\theta
D_{OL}}{u_{ps}}-1\bigg)}+\bar{b}+\mathcal{O}(u-u_{ps}), \label{alf1}
\end{eqnarray}
with
\begin{eqnarray}
&\bar{a}&=\frac{R(0,r_{ps})}{2\sqrt{q(r_{ps})}}, \nonumber\\
&\bar{b}&= -\pi+b_R+\bar{a}\log{\bigg\{\frac{2q(r_{ps})C(r_{ps})}{u_{ps}|A(r_{ps})|[D(r_{ps})+u_{ps}A(r_{ps})]}\bigg\}}, \nonumber\\
&b_R&=I_R(r_{ps}), \nonumber\\
&u_{ps}&=\frac{-D(r_{ps})+\sqrt{A(r_{ps})C(r_{ps})+D^2(r_{ps})}}{A(r_{ps})},\label{coa1}
\end{eqnarray}
where the quantity $D_{OL}$ is the distance between observer and
gravitational lens. Making use of Eqs. (\ref{alf1}) and (\ref{coa1}),
we can study the properties of strong gravitational lensing in the
rotating quasi-Kerr spacetime. In Fig. \ref{ coefficients}, we plotted
the changes of the coefficients $u_{ps}$, $\bar{a}$ and $\bar{b}$
with $a$ for a different quadrupolar
correction parameter $\xi$ . It is
shown that the coefficients ($\bar{a}$ and $\bar{b}$ ) in the strong
field limit are functions of the parameters $a$ and  $\xi$. We can see that $u_{ps}$ has a similar behavior as $r_{ps}$
 shown in Fig. \ref{figure1}. For a fixed $a$, the coefficient $\bar{a}$
decreases but coefficient $\bar{b}$ increases with increasing of the quadrupolar
correction parameter $\xi$. With the help of the coefficients $\bar{a}$ and $\bar{b}$  we plotted the change of the deflection angles evaluated
at $u=u_{ps}+0.00326$ with $\xi$ in Fig. \ref{figure3}. It is shown that in the strong field limit the deflection angles have the similar properties of the
coefficient $\bar{a}$ when the quadrupolar correction parameter $\xi$ takes a positive value.  Moreover, it is interesting to note that, when the
quadrupolar correction parameter $\xi$ takes a negative value,
the deflection angles can increase to a peak value for a certain $a$.
And the deflection angles of the quasi-Kerr compact object with the negative quadrupolar correction parameter $\xi$ is larger than the case of Kerr black hole. These results tell us that we could get the information of the compact object with quadrupole moment by means of strong gravitational lensing.

\subsection{Observable quantities of strong gravitational lensing}

Let us now to study the effect of the quadrupolar
correction parameter $\xi$
on the observable quantities of strong gravitational lensing. Here we
consider only the case in which the source, lens and observer are
highly aligned so that the lens equation in strong gravitational
lensing can be approximated well as \cite{Bozza3}
\begin{eqnarray}
\gamma=\frac{D_{OL}+D_{LS}}{D_{LS}}\theta-\alpha(\theta) \; mod
\;2\pi,
\end{eqnarray}
where $D_{LS}$ is the lens-source distance and $D_{OL}$ is the
observer-lens distance, $\gamma$ is the angle between the direction
\begin{figure}[ht]
\begin{center}
\includegraphics[scale=1.1]{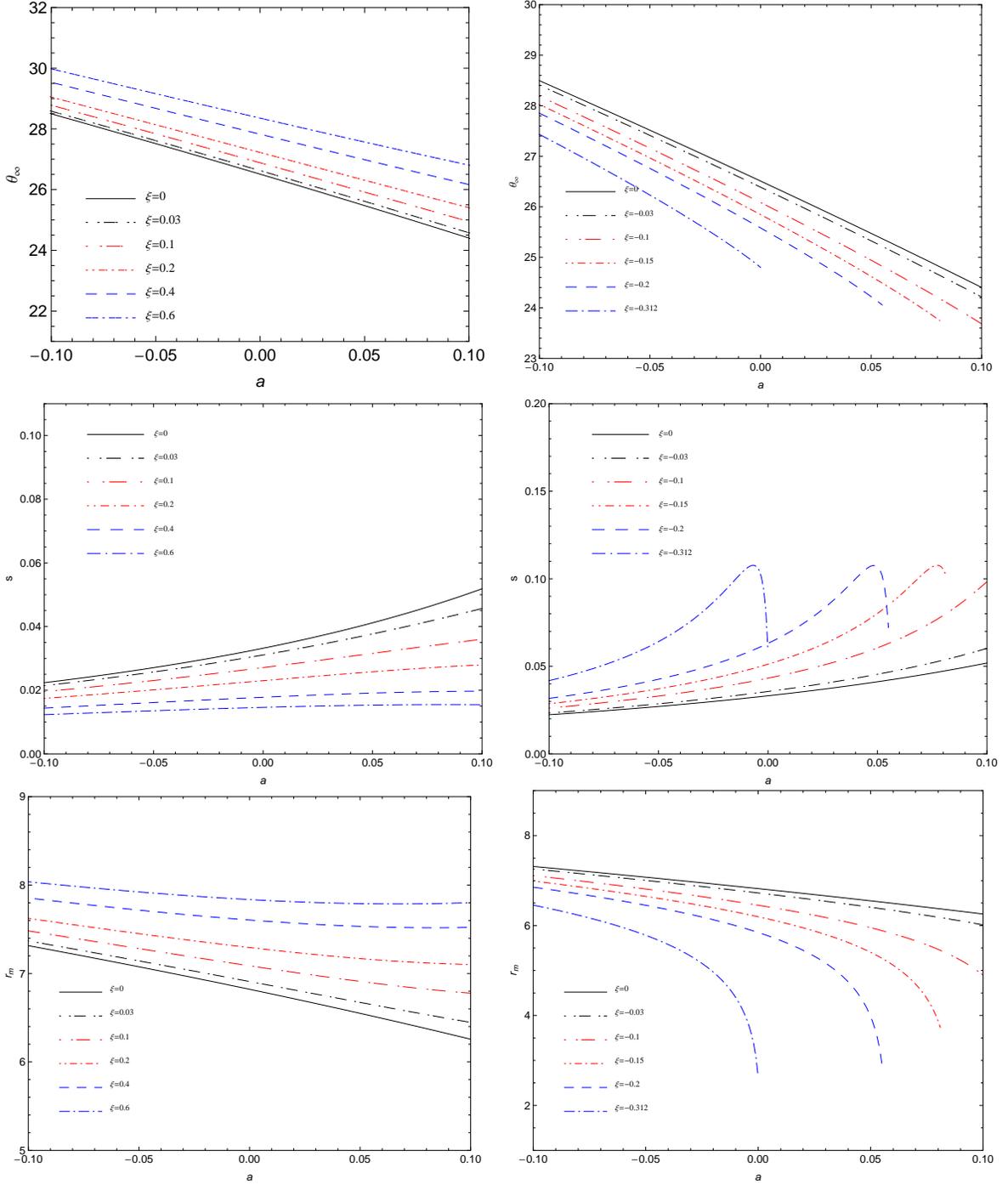}\caption{Variation of the innermost relativistic image $\theta_{\infty}$, the relative magnitudes $r_m$  and the angular separation
$s$ with the parameter $a$ for different $\xi$. Here, we set $2M=1$.}\label{figure4}
\end{center}
\end{figure}
\begin{table}
\begin{center}
\begin{tabular}{|c|c|c|c|c|c|c|c|c|}
 \hline &\multicolumn{8}{c|}{$\theta_{\infty}$($\mu$
arcsec)}\\
\hline $a$ &$\xi=-0.3$&$\xi=-0.2$&$\xi=-0.1$&$\xi=0$&$\xi=0.1$&$\xi=0.2$&$\xi=0.4$&$\xi=0.6$\\
\hline
 -0.10& 27.478&27.854&28.190&28.497&28.781& 29.047 &29.534&29.978\\
 \hline
-0.05&26.289&26.759&27.160&27.516&27.839&28.138&28.678&29.161\\
\hline
0& 24.905&25.581&26.085&26.510&26.885&27.224&27.827&28.356 \\
\hline
0.05& &24.222&24.943&25.474&25.919&26.310&26.987&27.569\\
 \hline
0.10& & &23.682&24.403&24.946&25.403&26.167&26.808
 \\
\hline\hline
 \hline &\multicolumn{8}{c|}{$s$($\mu$
arcsec)}\\
\hline $a$ &$\xi=-0.3$&$\xi=-0.2$&$\xi=-0.1$&$\xi=0$&$\xi=0.1$&$\xi=0.2$&$\xi=0.4$&$\xi=0.6$\\
\hline
 -0.10& 0.0405&0.0317&0.0262&0.0223&0.0196& 0.0174 &0.0144&0.0123\\
 \hline
-0.05&0.0610&0.0426&0.0330&0.0271&0.0230&0.0201&0.0161&0.0135\\
\hline
0& 0.1069&0.0631&0.0417&0.0332&0.0271&0.0223&0.0179&0.0147 \\
\hline
0.05& &0.1069&0.0606&0.0412&0.0316&0.0258&0.0191&0.0153\\
 \hline
0.10& & &0.0806&0.0519&0.0360&0.0279&0.0197&0.0154
 \\
 \hline\hline
 \hline &\multicolumn{8}{c|}{$r_m$(magnitudes)}\\
\hline $a$ &$\xi=-0.3$&$\xi=-0.2$&$\xi=-0.1$&$\xi=0$&$\xi=0.1$&$\xi=0.2$&$\xi=0.4$&$\xi=0.6$\\
\hline
 -0.10& 6.5018&6.8540&7.1120&7.3151&7.4821& 7.6236 &7.8543&8.0377\\
 \hline
-0.05&5.8734&6.4513&6.8132&7.0761&7.2820&7.4509&7.7174&7.9233\\
\hline
0& 3.9225&5.8458&6.4523&6.8219&7.0873&7.2940&7.6053&7.8366 \\
\hline
0.05& &3.9744&5.9528&6.5499&6.9108&7.1689&7.5333&7.7908\\
 \hline
0.10& & &4.8954&6.2569&6.7774&7.1022&7.5179&7.8018
 \\
 \hline\hline
\end{tabular}
\end{center}
 \caption{Numerical estimation for main observables in
the strong field limit for the black hole at the center of our
Galaxy, which is supposed to be described by in the slowly-rotating
black hole in the quasi-Kerr spacetime.}\label{tab11}
\end{table}
of the source and the optical axis, $\theta=u/D_{OL}$ is the angular
separation between the lens and the image. Following
ref. \cite{Bozza3}, we can find that the angular separation between
the lens and the n-th relativistic image is
\begin{eqnarray}
\theta_n\simeq\theta^0_n\bigg(1-\frac{u_{ps}e_n(D_{OL}+D_{LS})}{\bar{a}D_{OL}D_{LS}}\bigg),
\end{eqnarray}
with
\begin{eqnarray}
\theta^0_n=\frac{u_{ps}}{D_{OL}}(1+e_n),\;\;\;\;\;\;e_{n}=e^{\frac{\bar{b}+|\gamma|-2\pi
n}{\bar{a}}},
\end{eqnarray}
where the quantity $\theta^0_n$ is the image positions corresponding to
$\alpha=2n\pi$, and $n$ is an integer. According to the past
oriented light ray which starts from the observer and finishes at
the source the resulting images stand on the eastern side of the
black hole for direct photons ($a>0$) and are described by positive
$\gamma$. Retrograde photons ($a<0$) have images on the western side
of the compact object and are described by negative values of $\gamma$.
In the limit $n\rightarrow \infty$, we can find that
$e_n\rightarrow 0$, which means that the relation between the
minimum impact parameter $u_{ps}$ and the asymptotic position of a
set of images $\theta_{\infty}$ can be simplified further as
\begin{eqnarray}
u_{ps}=D_{OL}\theta_{\infty}.\label{uhs1}
\end{eqnarray}
In order to obtain the coefficients $\bar{a}$ and $\bar{b}$, we
needs to separate at least the outermost image from all the others.
As in refs. \cite{Bozza2,Bozza3},  we consider here the simplest case
in which only the outermost image $\theta_1$ is resolved as a single
image and all the remaining ones are packed together at
$\theta_{\infty}$. Thus the angular separation between the first
image and other ones can be expressed as
\cite{Bozza2,Bozza3,Gyulchev1}
\begin{eqnarray}
s=\theta_1-\theta_{\infty}=\theta_{\infty} e^{\frac{\bar{b}-2\pi}{\bar{a}}}.\label{ss1}
\end{eqnarray}
By measuring $s$ and $\theta_{\infty}$, we can obtain the
strong deflection limit coefficients $\bar{a}$, $\bar{b}$ and the
minimum impact parameter $u_{ps}$. Comparing their values with those
predicted by the theoretical models, we can obtain information of the compact object.

The mass of the central object of our Galaxy is estimated recently
to be $4.4\times 10^6M_{\odot}$ \cite{Genzel1} and its distance is
around $8.5kpc$, so that the ratio of the mass to the distance
$M/D_{OL} \approx2.4734\times10^{-11}$.  Making use of Eqs.
(\ref{coa1}), (\ref{uhs1}) and  (\ref{ss1})  we can estimate the
values of the coefficients and observable quantities for gravitational lens
in the strong field limit. The numerical value for the angular
position of the relativistic images $\theta_{\infty}$, the
angular separation $s$ and the relative magnitudes $r_m$ are plotted in Fig. \ref{figure4} and listed in table \ref{tab11}. We find that with the increase of $\xi$, the
angular position of the relativistic images $\theta_{\infty}$
decreases for both the direct photons $(a>0)$ and the
retrograde photons $(a<0)$. The variation of the angular
separation $s$ and the relative magnitudes $r_m$ with $\xi$ is similar to that of the deflection angle $\alpha(\theta)$ and the deflection angle coefficient $\bar{b}$, respectively.

Now, we have found that the numerical value for the angular position
of the innermost relativistic images $\theta_{\infty}$ of quasi-Kerr
compact object is $23-29 \mu arcsec$. In principle, such a optical
resolution is reachable by very long baseline interferometry (VLBI)
projects, but we may be aware that relativistic images are difficult
to detect and the resolution of the most powerful modern instruments
is currently insufficient to perform such high precision astrometry.
However, innovative near-infrared interferometry instruments are now
under development at the very large Telescope and Keck, i.e.,
advanced radio interferometry between space and Earth (ARISE). These
instruments have been conceived to achieve an astrometric accuracy
of $10-100 \mu arcsec$ in combination with milli-arcsec
angular-resolution imaging \cite{vbozza3}. Thus we can observe
relativistic images within a not so far future.

The relativistic images obtained by means of strong gravitational lensing in our paper is different from the
the images of the quasi-Kerr spacetime  obtained by the use of a ray-tracing algorithm in ref. \cite{Johannsen}.  However, Johannsen \textit{et al} showed that images of the quasi-Kerr compact object become oblate or prolate depending on the sign and value of the parameter $\xi$, which has a link with the properties of the minimum impact parameter $u_{ps}$ and the radius of photon sphere $r_{ps}$ in our paper, i.e.,  when the parameter $\xi$ takes the positive (negative) value, the photon-sphere
radius $r_{ps}$ and the minimum impact parameter $u_{ps}$ are larger (smaller) than
the results obtained in the Kerr black hole.
 Johannsen \textit{et al} use the degree of asymmetry and displacement of the photon ring as a direct measure of the violation of the no-hair theorem, while we use observables (the angular
position of the relativistic images $\theta_{\infty}$, the
angular separation $s$ and the relative magnitudes $r_m$) to measure the relativistic images. It is interesting to note the angular size of the diameter of the photon ring $\theta_{ring}=53\mu arcsec$ of a slowly rotating Kerr black hole obtained in ref. \cite{Johannsen} is approximatively equal to the size of the angular diameter of
the innermost relativistic images $\theta_{\infty}$ for the case
$a=0, \xi=0$ in table \ref{tab11} of our paper (noted that the data listed  in table \ref{tab11} is the radius of  $\theta_{\infty}$). Thus, the observation of both the images of black hole in ref. \cite{Johannsen} and the relativistic
images caused by strong gravitational lensing in our paper
require astronomical instruments have high optical resolution. Based on the above discussion
we can conclude that, with the electromagnetic spectrum
proposed in ref. \cite{Johannsen} and the relativistic images of our paper, we can
comprehensively observer how astronomical compact object deviates from the Kerr black hole.

\section{summary}

In this paper, in order to test how astronomical compact object deviates from the Kerr black hole, we investigate the features of light propagation on the equatorial plane of the rotating quasi-Kerr spacetime proposed
by Glampedakis \textit{et al} \cite{Glampedakis}. Assuming
that the massive compact object at the center of our galaxy can be described by quasi-Kerr spacetime,
we obtain the photon radius, the coefficients and observable quantities of the strong gravitational lensing.
We find that the photon-sphere radius exist only when the quadrupolar correction parameter takes the value $\xi>-0.302$ for the photons in the prograde orbit $(a>0)$. Moreover, as the quadrupolar correction parameter $\xi$ becomes negative the radius of the photon sphere becomes smaller,  which implies that the photons are more easily captured by the quasi-Kerr compact object with the negative quadrupolar correction parameter $\xi$ than that of the Kerr black hole. It is interesting to note that, when the quadrupolar correction parameter $\xi$ takes the positive (negative) value, the photon-sphere radius $r_{ps}$, the minimum impact parameter $u_{ps}$, the coefficient $\bar{b}$, the relative magnitudes $r_m$ and the angular position of the relativistic images $\theta_{\infty}$ are larger (smaller)  than the results obtained in the Kerr black hole, but the coefficient $\bar{a}$, the deflection angle $\alpha(\theta)$ and the angular separation $s$ are smaller (larger) than that of the Kerr black hole. Based on the above results, we come to the conclusion that there are some significant effects of the quadrupolar correction parameter $\xi$ on
the coefficients and observable parameters of the strong gravitational lensing.  These results, in principle, may provide a possibility to test how astronomical black holes with arbitrary quadrupole moments deviate from the Kerr black hole in the future astronomical observations.

\section{\bf Acknowledgments}

This work was supported by the National Natural Science  Foundation
of China under Grant No. 11175065, 10935013; the National Basic
Research of China under Grant No. 2010CB833004; the SRFDP under Grant
No. 20114306110003; PCSIRT, No. IRT0964;
the Hunan Provincial Natural Science Foundation of China under Grant
No 11JJ7001; Hunan Provincial Innovation Foundation For Postgraduate
under Grant No CX2011B185; and Construct Program of the National
Key Discipline.

\end{document}